\documentclass[floatfix,lengthcheck,showpacs,amssymb,amsmath,amsfonts,twocolumn]{aastex62}

\usepackage{lineno}
 \usepackage[utf8]{inputenc}
\usepackage{color}
\usepackage{graphicx}
\usepackage{float}
\usepackage{subfigure}
\usepackage{amsmath}
\usepackage{acronym}
\usepackage{multirow}
\usepackage{tabu}
\usepackage{tikz}
\usetikzlibrary{arrows,shapes,trees,decorations.pathreplacing}
\usepackage{import}
\usepackage{svn-multi}
\usepackage{lineno}
\usepackage{gensymb}
\hypersetup{
    colorlinks=true,
    linkcolor=blue,
    filecolor=magenta,
    urlcolor=cyan,
}

\newcommand{\chieff}{\ensuremath{\chi_{\mathrm{eff}}}}

\newcommand{\msun}{\ensuremath{\mathrm{M}_{\odot}}}

\newcommand{\measmchirp}{\ensuremath{1.30^{+0.02}_{-0.03}}}
\newcommand{\measdist}{\ensuremath{187^{+99}_{-87}}}
\newcommand{\measconedist}{\ensuremath{224^{+88}_{-78}}}

\begin{document}
\title[]{Potential Gravitational-wave and Gamma-ray Multi-messenger Candidate from Oct. 30, 2015}

\correspondingauthor{Alexander H. Nitz}
\email{alex.nitz@aei.mpg.de}

\author[0000-0002-1850-4587]{Alexander H. Nitz}
\affil{Max-Planck-Institut f{\"u}r Gravitationsphysik (Albert-Einstein-Institut), D-30167 Hannover, Germany}
\affil{Leibniz Universit{\"a}t Hannover, D-30167 Hannover, Germany}

\author[0000-0002-1850-4587]{Alex B. Nielsen}
\affil{Max-Planck-Institut f{\"u}r Gravitationsphysik (Albert-Einstein-Institut), D-30167 Hannover, Germany}
\affil{Leibniz Universit{\"a}t Hannover, D-30167 Hannover, Germany}

\author[0000-0002-0355-5998]{Collin D. Capano}
\affil{Max-Planck-Institut f{\"u}r Gravitationsphysik (Albert-Einstein-Institut), D-30167 Hannover, Germany}
\affil{Leibniz Universit{\"a}t Hannover, D-30167 Hannover, Germany}

\keywords{gamma ray bursts --- gravitational waves --- stars: neutron}

\begin{abstract}
We present a search for binary neutron star mergers that produced gravitational-waves during the first observing run of Advanced LIGO and gamma-ray emission seen by either \textit{Swift}-BAT or \textit{Fermi}-GBM, similar to GW170817 and GRB 170817A. We introduce a new method using a combined ranking statistic to detect sources that do not produce significant gravitational-wave or gamma-ray burst candidates individually. The current version of this search can increase by 70\% the detections of joint gravitational-wave and gamma-ray signals. We find one possible candidate observed by LIGO and \textit{Fermi}-GBM, 1-OGC 151030, at a false alarm rate of 1 in 13 years. If astrophysical, this candidate would correspond to a merger at $\measdist\,$Mpc with source-frame chirp mass of $\measmchirp\,\msun$. If we assume the viewing angle must be $<30^{\circ}$ to be observed by \textit{Fermi}-GBM, our estimate of the distance would become $\measconedist\,$Mpc. By comparing the rate of binary neutron star mergers to our search-estimated rate of false alarms, we estimate that there is a 1 in 4 chance this candidate is astrophysical in origin.
\end{abstract}

\section{Introduction}
\label{sec:intro}

The detection of the binary neutron star (BNS) coalescence GW170817 \citep{TheLIGOScientific:2017qsa} was a watershed moment in astronomy. In a triumph for multi-messenger astronomy, gravitational-wave (GW) observations turned what would otherwise have been a relatively unremarkable Gamma-ray Burst (GRB) detection by the \textit{Fermi}-Gamma-ray Burst Monitor (GBM), into one of the most studied astronomical transients of all time \citep{Goldstein:2017mmi, GBM:2017lvd}.
The observation of GW170817 immediately raised the question of whether other coincident detections were possible in archival data that had previously been missed. The success of GW170817 provided greater constraints on what should be searched for, demonstrating the importance of quieter, potentially off-axis GRBs \citep{Burns:2018dbd, vonKienlin:2019otj, Mooley:2018clx}.

As a gravitational-wave event, GW170817 was very loud and would not have been missed even without the multi-messenger electromagnetic observations \citep{TheLIGOScientific:2017qsa}. However, multi-messenger astronomy provides the possibility of combining data sets from different observatories to promote events that would not have been convincing on their own into solid astrophysical candidates, for example the TeV-energy neutrino IceCube-170922A coincident with a gamma-ray flare from a blazar \citep{IceCube:2018dnn}. This is because a coincident search between two observation sets can substantially reduce the background that exists in either set independently.

In this paper we develop a method to perform such coincident multi-messenger searches using gravitational-wave and gamma-ray data. We demonstrate its performance on open archival data from Advanced LIGO's first observing run (O1) and coincident \textit{Fermi}-GBM data. We specifically target sub-threshold candidates that would not be significant in either individual data set, but also include autonomously detected GRBs. We restrict our search to BNS systems similar to the already observed GW170817. Previous searches of these data sets have either looked for weak GW signals associated with autonomously detected GRBs \citep{Abbott:2016cjt} or have used sub-threshold GW candidates without imposing the constraint that they should be BNS systems \citep{Burns:2018pcl}.

 The LIGO Scientific and Virgo Collaborations have released a list of candidate triggers with a false alarm rate less than 1 per 30 days~\citep{LIGOScientific:2018mvr}. Within this set, most confident GW detections are seen at false alarm rates less than one per hundred years. Over the past decade, GBM has provided thousands of autonomously generated triggers \citep{Goldstein:2017mmi}. During O1, 42 gamma-ray bursts were seen by a number of different detectors \citep{Hurley:2002wv, Gruber:2014iza, Lien:2016zny}, with 15 of these being the short/hard type believed to be associated with BNS mergers~\citep{Abbott:2016cjt}.
 Nearby short GRBs may not be especially luminous~\citep{Burns:2015fol, Monitor:2017mdv}. Although GW detectors do not currently have the range needed to observe all GRBs, the observation of GW170817 has raised the importance of studying nearby and potentially less energetic GRBs. GW170817 was observed to be between two and six orders of magnitude less energetic than other short GRBs~\citep{Monitor:2017mdv}.

Based on the estimated viewing angle of GW170817, it is reasonable to assume that BNSs produce GRBs that are beamed and may be observable within a cone of $\sim30\degree$~\citep{Schutz:2011tw, Fong:2015oha, Finstad:2018wid}. With this restriction on inclination, we estimate the sky-averaged sensitive distance of the GW search to be 140 Mpc during the time both LIGO detectors were observing and 100 Mpc in single-detector observing time at a false alarm rate of 1 per year. Using an approximate rate of BNS mergers of $1000\,\mathrm{Gpc}^{-3}\mathrm{yr}^{-1}$ \citep{LIGOScientific:2018mvr}, the expected number of BNSs in this volume during O1 is $\sim 1.98$ of which we expect $0.27$ events to be beamed towards Earth.  Our joint GW-GRB search finds one potential coincident candidate at a false alarm rate of 1 in 13 years and source-frame chirp mass $\sim1.30\,\msun$. By comparing the expected number of signals to the search-estimated number of false alarms, we estimate that this candidate has a 1 in 4 chance of being astrophysical.

\section{Combined Search for GW-GRB Coincidences}

We search for multi-messenger GW-GRB candidates by correlating sub-threshold GW candidates from the public 1-OGC catalog~\citep{Nitz:2018imz} including single-detector GW candidates with GRB candidates from the combined set of public \textit{Swift}-BAT~\citep{Lien:2016zny} and \textit{Fermi}-GBM candidates~\citep{Gruber:2014iza, vonKienlin:2014nza, Bhat:2016odd}, as well as our own set of sub-threshold candidates derived from a straightforward analysis of the archival \textit{Fermi}-GBM data. This analysis is targeted at finding GW-GRB coincidences produced by BNS mergers similar to GW170817 and GRB 170817A. Using a simulated set of gravitational-wave signals, we estimate that, relative to a GW search alone, this analysis is able to achieve a $\sim20\%$ improvement in sensitive distance. This is comparable to the coherent follow-up of GRBs in~\cite{Abbott:2016cjt,Williamson:2014wma}.

\begin{figure}[]
  \centering
    \includegraphics[width=\columnwidth]{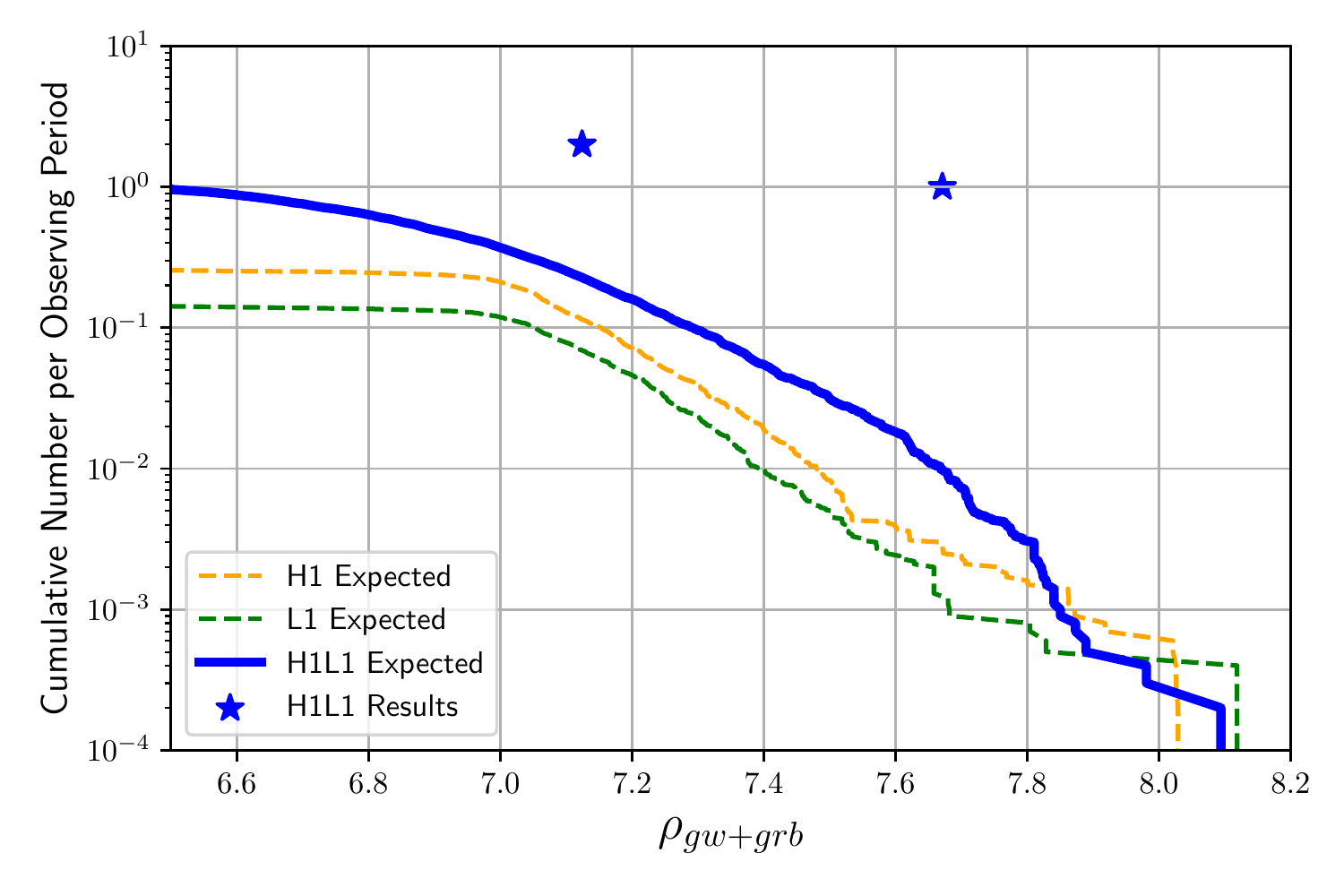}
\caption{The results from our search for GW-GRB coincidences. The lines show the
cumulative number of expected accidental coincidences during times when only LIGO-Hanford (orange), only LIGO-Livingston (green), and both LIGO instruments (blue) are observing as a function of our ranking statistic $\rho_{gw-grb}$. No candidate
events were found during times when a single LIGO detector was observing. The two candidates found when both LIGO instruments were observing are shown with blue stars. The most significant is 1-OGC 151030. We expect an accidental candidate at least as significant as 1-OGC 151030 once in 100 48-day analyses. This corresponds to a false alarm rate of 1 in 13 years.}
\label{fig:result}
\end{figure}

\subsection{Gravitational-wave Candidates}
 The 1-OGC catalog~\citep{Nitz:2018imz} contains candidate mergers during times when the LIGO instruments were both observing, and is particularly suitable for multi-messenger follow-up due to the low threshold for candidate inclusion. We select candidates from this catalog that are consistent with the expected population of BNSs (as discussed below) and use the public LIGO data~\citep{Vallisneri:2014vxa} to produce sky localizations for each GW candidate. In addition to candidate events from the 1-OGC catalog, which only includes candidates when both LIGO instruments were observing, we also search times when only one of the two LIGO instruments (Hanford or Livingston) was observing. While single-detector GW observations are typically difficult to confidently claim on their own, they can be confirmed by a GRB counterpart. Coincident LIGO observing time accounts for $\sim48$ days of data; single-detector observing time adds an additional $\sim44$ days.

Considering astrophysical observations, including GW170817, we target BNSs with component masses $1.33\pm 0.09 M_{\odot}$~\citep{Ozel:2016oaf}. The 1-OGC catalog records the detector-frame component masses $m_{1,2}$ and aligned dimensionless spin components $\chi_{1z,2z} = S_{1z,2z}/m_{1,2}^2$ of the gravitational-wave template waveform associated with each candidate event. We select candidates consistent with this population by placing constraints on the chirp mass $\mathcal{M} = (m_1m_2)^{3/5} / (m_1+m_2)^{1/5}$ and effective spin $\chieff = (\chi_{1z} m_1 + \chi_{2z} m_2)/(m_1+m_2)$ \citep{Ajith:2011ec}. These parameters are more accurately measured than the component masses and spins directly~ \citep{Cutler:1994ys, Ohme:2013nsa}. We find the detector-frame constraints $1.03<\mathcal{M}<1.36$ and $|\chieff| < 0.2$ are effective at recovering a simulated population of sources and allows for a $2\sigma$ deviation in the masses. This range also accounts for the average shift in the observed masses due to cosmological redshift.

Each candidate event from the 1-OGC sample was observed by both LIGO detectors and already has an assigned ranking statistic $\tilde{\rho}_c$ which is proportional to signal-to-noise (SNR) under ideal conditions and inversely proportional to the luminosity distance~\citep{Nitz:2017svb,Nitz:2018imz}. There are 3395 1-OGC candidates with $\tilde{\rho}_c > 6.5$ selected by the cuts described in the previous paragraph. We use PyCBC Inference~\citep{Biwer:2018osg,pycbc-github} to estimate their sky localizations. We fix the component masses and spins of the source binary to those found in the 1-OGC catalog. This is a reasonable approximation since the estimation of the sky localization is largely independent from the other parameters~\citep{Singer:2015ema}. For ~$3\%$ of these candidates the sky-location estimation failed to converge and instead we use the combined Hanford and Livingston detector response, which is the sensitivity of the detectors to a given sky location. Closer inspection of many of these cases indicated that they lay in stretches of time with non-stationary data.

For single-detector candidates, we select triggers that lie in our chosen parameter space, are the loudest candidate within 10 seconds, and that have re-weighted SNR $\tilde{\rho}_{H,L} > 7$~\citep{Babak:2012zx,Nitz:2018imz}. The sky localization assigned for single-detector candidates is the detector response of its respective LIGO observatory.

\begin{figure}[h]
  \centering
    \includegraphics[width=\columnwidth]{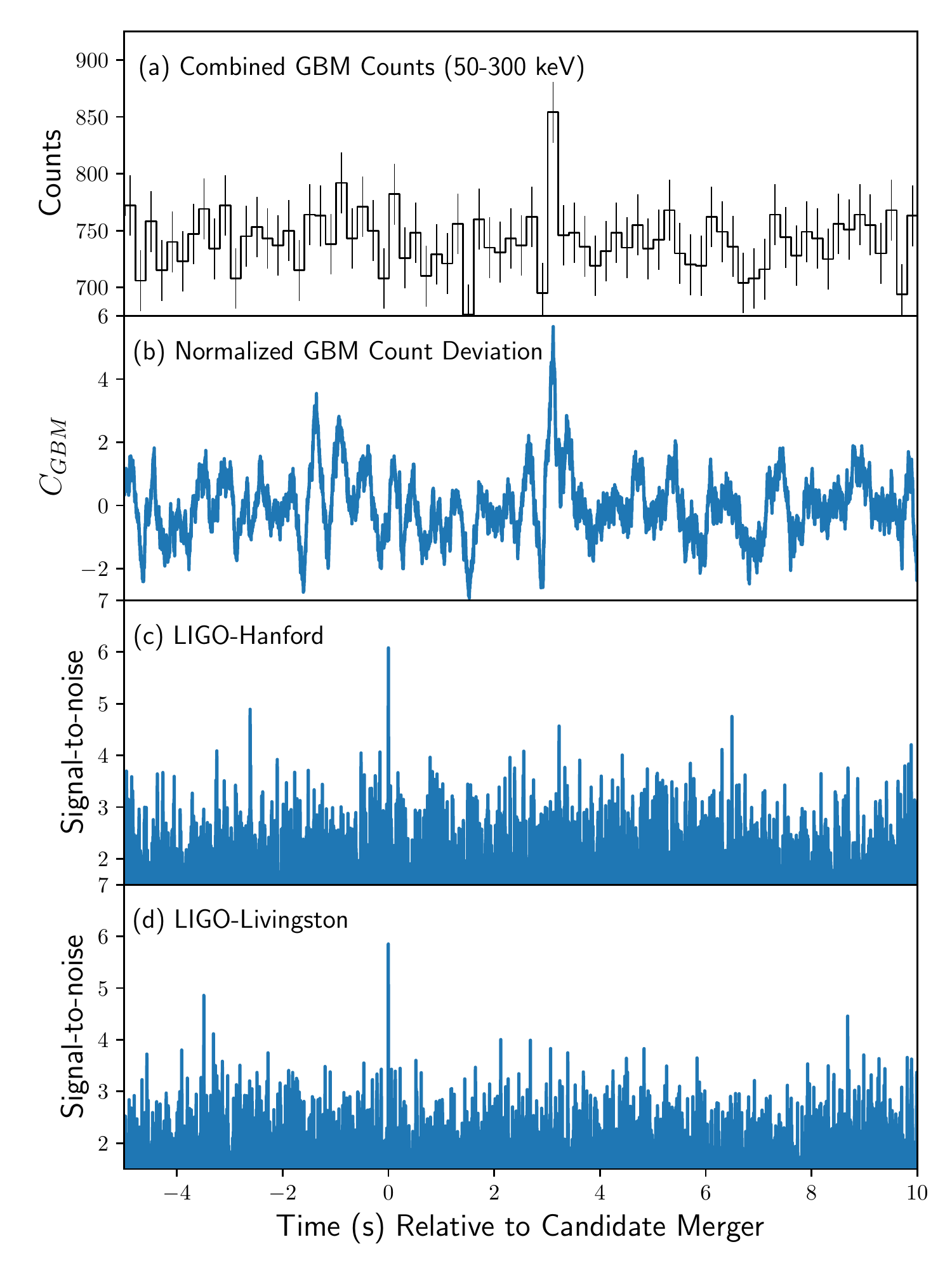}
\caption{Time series results from the \textit{Fermi}-GBM and the LIGO observatories at the time
of 1-OGC 151030. (a) shows a simple histogram of the GBM data from all 12 NaI detectors with a bin width of 0.2s. (b) shows the normalized statistic $C_{GBM}$ that we use to determine the presence of a gamma-ray excess. We impose a threshold of $C_{GBM} > 5.5$ to determine our population of low significance samples. (c) and (d) show the signal-to-noise time series for the gravitational-wave template waveform used to find this candidate. Two peaks in the gravitational-wave SNR are visible at nearly the same time followed by a small peak in the normalized deviation in GBM counts 3.1 seconds later.}
\label{fig:snrgbm}
\end{figure}

\begin{figure*}[]
  \centering
    \includegraphics[width=\textwidth]{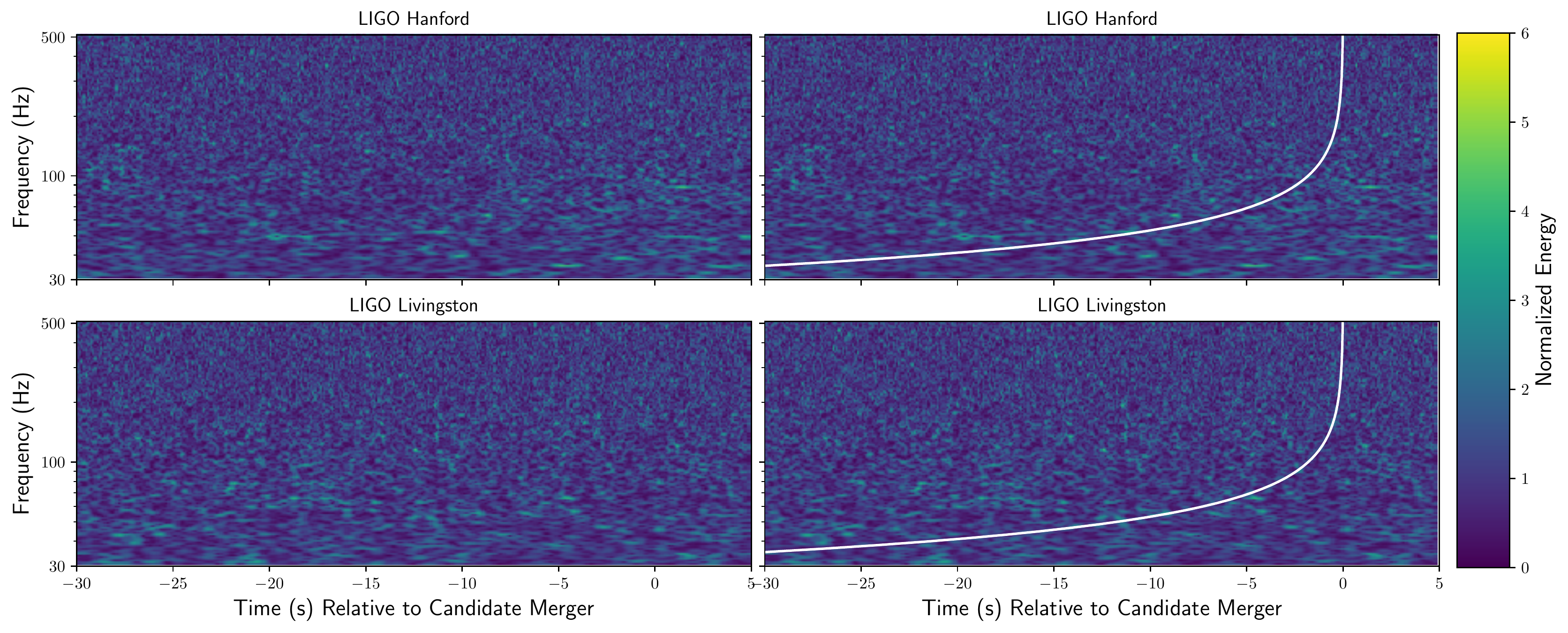}
\caption{Time-frequency representations of the GW data using the Q-transform (Q=100) around the time of the 1-OGC 151030 candidate. Upper (lower) panels show data from the LIGO Hanford (Livingston) observatory. No loud transient noise similar to that observed during GW170817 is visible. The left panels show the raw data alone. The right panels overlay the same data with a track of a gravitational-wave signal with the parameters of 1-OGC 151030 to guide the eye. No visible track is expected in the raw data for quiet signals.}
\label{fig:data}
\end{figure*}

\subsection{Gamma-ray Burst Candidates}

 We generate our sample of GRBs separately from our set of gravitational-wave candidates. We include all short GRBs ($T_{90} < 2$s) reported by both \textit{Swift}-Burst Alert Telescope (BAT) and \textit{Fermi}-GBM from Sept. 2015 through Jan. 2016~\citep{Lien:2016zny,Gruber:2014iza, vonKienlin:2014nza, Bhat:2016odd}. We also include sub-threshold events reported by \textit{Fermi}-GBM during this time\footnote{\url{https://gammaray.nsstc.nasa.gov/gbm/science/sgrb_search.html}}. Many of these candidates have been previously searched and found to have no identifiable LIGO counterpart~\citep{Abbott:2016cjt}. Since the expected GRB luminosity for a given GW candidate is not well constrained, and indeed subluminous bursts are of interest, we also conduct our own search of \textit{Fermi}-GBM data with lowered thresholds. The aim is to include lower luminosity sources at the expense of the purity of the sample set.

We use archival time-tagged event (TTE) data from \textit{Fermi}-GBM which consists of discrete events with a time and energy range from each of the 12 NaI and 2 BGO detectors~\citep{2009ApJ...702..791M}. GRBs similar to GW170817 are bright in the range 50-300 keV~\citep{vonKienlin:2019otj} so we combine the recorded photon counts within this energy range from all 12 NaI detectors. We do not include the BGO detectors as they focus on a higher energy range. For each event we calculate the combined number of photon counts within a $\pm 0.1$s window. Since the overall observed count flux can vary significantly over tens of seconds, we normalize our results by subtracting out the locally measured mean and dividing by the standard deviation. To measure the local mean and standard deviation, we limit to times within $\pm 4$s centered around each event and exclude time within $\pm 0.5$s to prevent a candidate GRB from biasing the estimates. We finally threshold on this normalized count excess, $C_{GBM}$.

We find thresholding at $C_{GBM} > 5.5$ includes all previously identified \textit{Fermi} short GRBs in our sample set. At this threshold we identify $\sim1$ candidate GRB every 3 hours. The aim of this threshold is to minimize false negatives while preserving the statistical power of the analysis and so it necessarily reduces the purity of our GRB sample. For comparison, the lowest value of $C_{GBM}$ assigned to a previously identified GRB is $\sim 7.8$ and GRB 170817 A would have been $\sim 10.5$.

A sky localization is assigned to to each candidate GRB. For any event which was previously reported, we use the published value and uncertainty. It is left to future work to measure detailed sky localizations for candidate GRBs identified with our sub-threshold analysis. In this analysis we assign them an isotropic sky localization. In all cases, we exclude directions that are occulted by the Earth from the perspective of \textit{Fermi}.

\begin{figure*}
    \centering
    \includegraphics[width=0.32\textwidth]{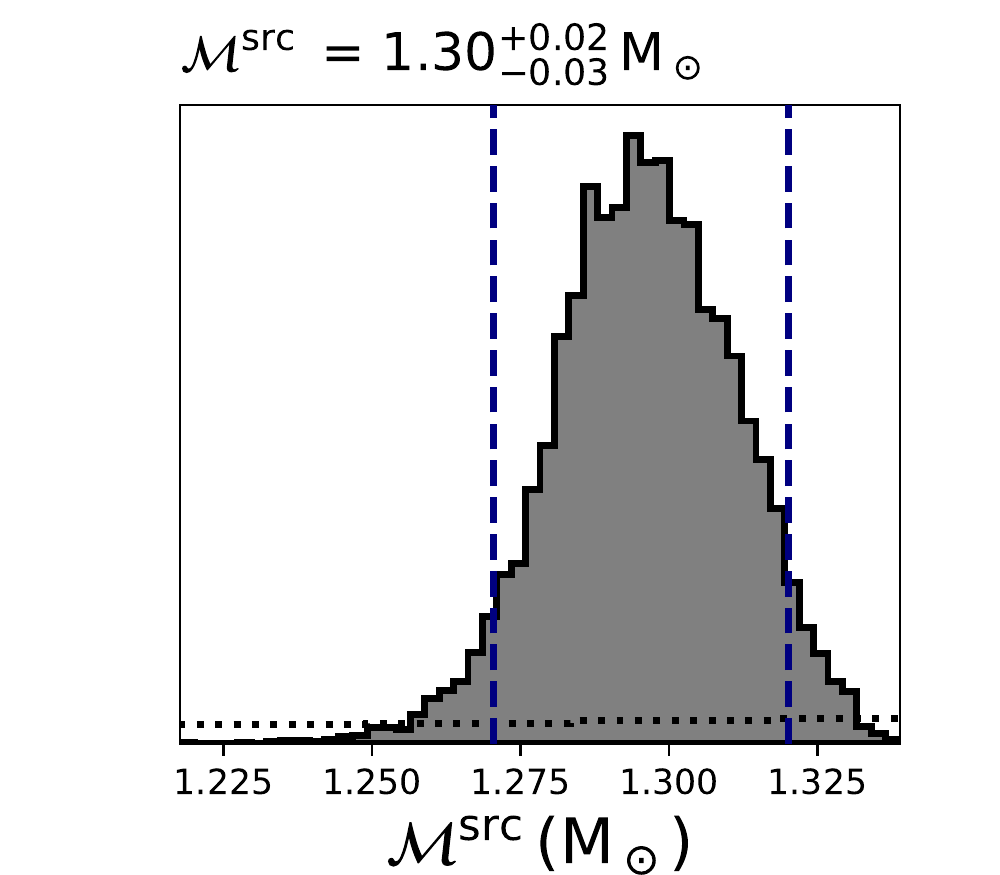}
    \includegraphics[width=0.32\textwidth]{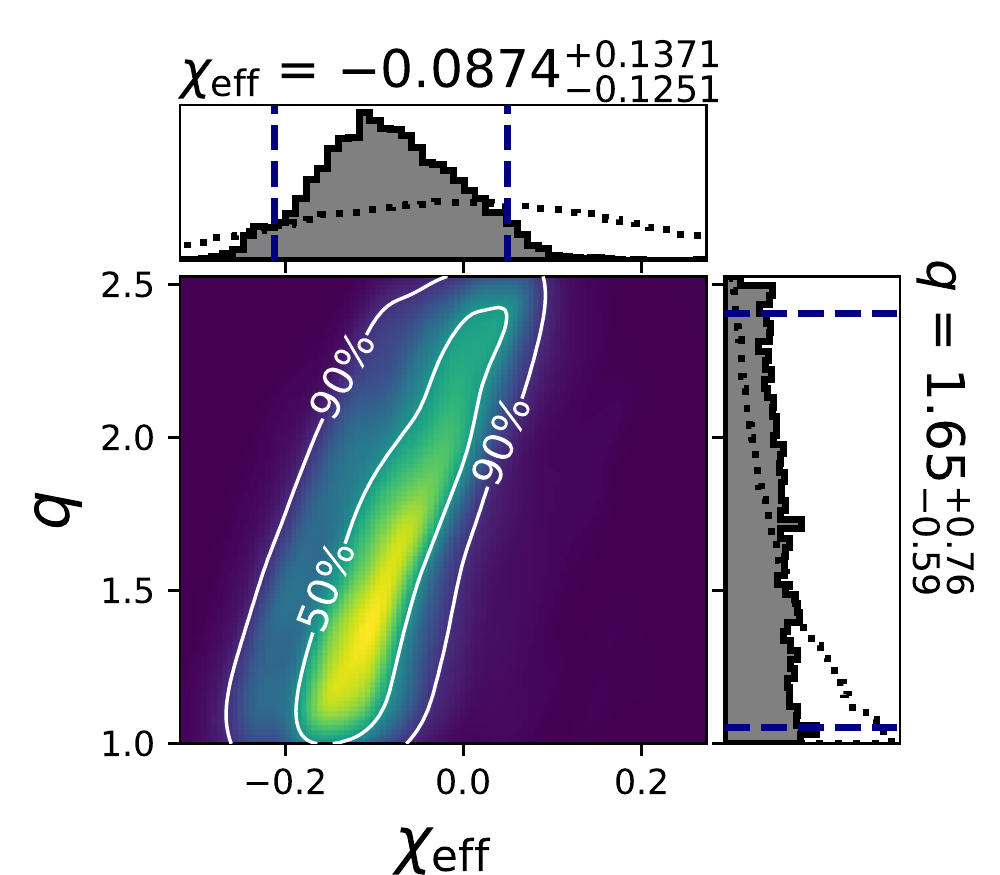}
    \includegraphics[width=0.32\textwidth]{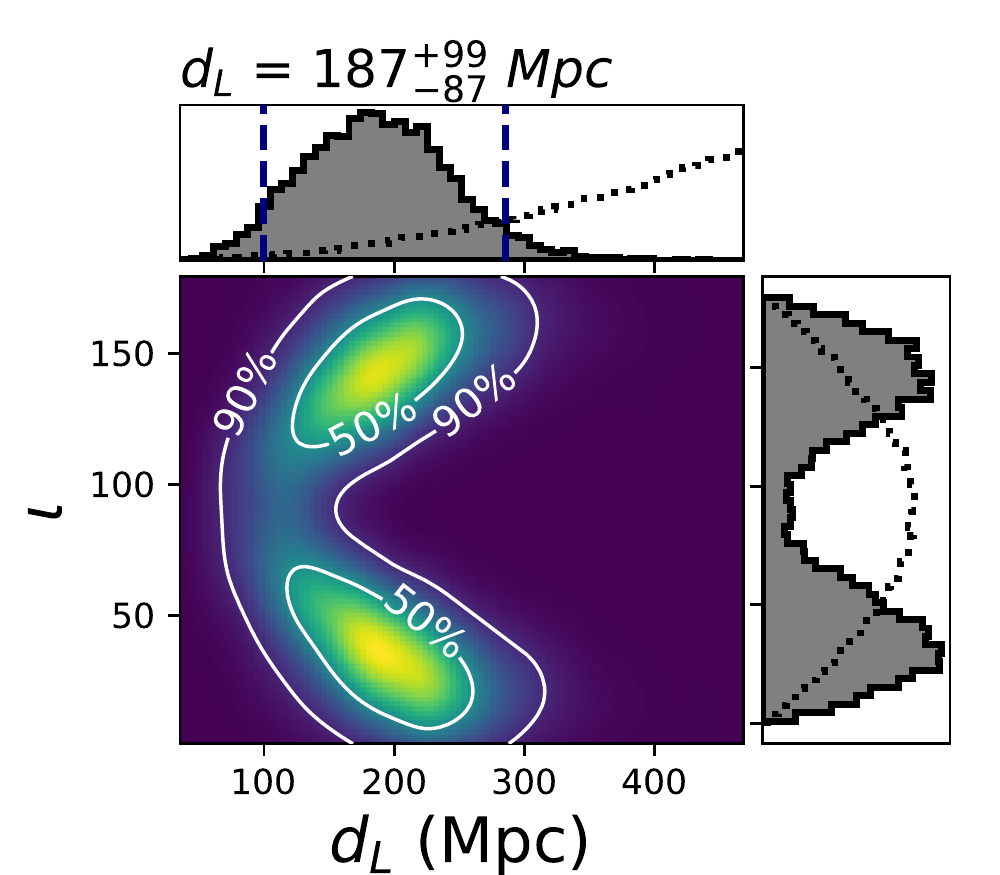}
    \caption{Marginal posterior distributions of the source chirp mass $\mathcal{M}^{\mathrm{src}}$
    (left), mass ratio $q$ (which we define to be $\geq 1$) and effective spin $\chieff$ (middle),
    and inclination $\iota$ and luminosity distance $d_L$ (right).
    The vertical, dashed blue lines show the 5 and 95 percentiles of the 1D marginal posteriors. The dotted black lines show the priors. The source-frame chirp mass is determined from the observed (detector-frame) chirp mass by $\mathcal{M}/(1+z)$, where the redshift $z$ is determined from $d_L$ using a standard $\Lambda$CDM cosmology \citep{Ade:2015xua}.}
    \label{fig:posteriors}
\end{figure*}

\subsection{Combining Gravitational and Gamma-ray Candidates}

 We look for temporal coincidences between our two independent GW and GRB candidate sets. We allow a GRB to occur 0-3.4s after the GW merger time (which happens to be broadly consistent with previous searches \citep{Abadie:2010uf}). Since we do not expect a GRB to arrive before an associated GW, this sets the lower bound on the time window. Astrophysical models are insufficient to constrain the upper time window, so we choose to use a symmetrical $\pm 1.7s$ window around the 1.7s time delay between GW170817 and GRB 170817A. Given the rate of GW and GRB candidates in our sample, this window corresponds to expecting one GRB-GW candidate by accidental coincidence in the 48 days when both LIGO instruments were observing.

We rank candidates according to their gravitational-wave likelihood and the agreement between the GW and GRB sky localization. This takes the form
\begin{equation}
\rho_{gw+grb}^2 = \rho_{gw}^2 + 2\log(B_{loc})
\end{equation}
where $\rho_{gw}$ is the SNR-like statistic associated with a given GW candidate event and $B_{loc}$ is the spatial Bayes factor employed by \cite{Ashton:2017ykh} which measures the agreement between the GW and GRB sky localizations. We do not include an explicit ranking of GRB candidates based on their flux due to the uncertainty in the relation between GW amplitude and GRB flux.

Under the assumption that the non-astrophysical backgrounds of GRB and GW detectors are not correlated in time, we determine the background of accidental coincidences by shifting the GRB candidates in time. We create $10^4$ time-shifted analyses, which allows us to measure the false alarm rate as a function of our ranking statistic $\rho_{gw+grb}$. We estimate separate backgrounds for time when a single LIGO detector is operating and when both LIGO instruments are observing.

\section{Observational Results}
Figure~\ref{fig:result} summarizes the results of our analysis. We find no candidate events during times when only a single LIGO instrument was observing, and two candidate events during times when both instruments were observing. The loudest candidate was observed at a false alarm rate of 1 per 13 years. Taking into account the $\sim92$ days of data analyzed, which includes single-detector operation, this event has a statistical significance of $2.0\sigma$. The second candidate is observed at a lower significance, and is consistent with the expected accidental coincident rate.

The candidate event, 1-OGC 151030, occurred on October 30, 2015 at 6:41:53 UTC (GPS time 1130222530). The event was recovered with SNR 5.7 in Livingston and 2.4ms later with SNR 6.0 in Hanford. This was followed by a peak of $C_{GBM} \sim 5.7$ in the observed gamma-ray counts 3.1s later. The observed counts from \textit{Fermi}-GBM along with the SNR time series from the GW template which recovered this candidate are shown in Fig.~\ref{fig:snrgbm}.

We estimate the probability that 1-OGC 151030 is astrophysical in origin by comparing the expected number of signal and noise events with similar $\rho_{gw+grb}$. We assume the rate of BNS mergers is $1000\,\mathrm{Gpc}^{-3}\mathrm{yr}^-1$, consistent with~\cite{LIGOScientific:2018mvr}, and that the sources are uniformly distributed in volume. The fraction of sources that are detectable as a function of $\rho_{gw+grb}$ is determined by searching for simulated gravitational-wave signals. We restrict our search to the portion of this population that could plausibly be visible with \textit{Fermi}-GBM by constraining the source inclination of our simulated population to be less than $30^{\circ}$ and taking into account that nearly half of potential sources may be missed due to detector downtime and blockage by the Earth. With these considerations, we count the number of signals that would be detected with $\rho_{gw+grb}$ within $\pm0.1$ of 1-OGC 151030 and compare to the number of background samples obtained by the search in the same range. We find the expected number is $\sim0.018$ and $\sim0.0064$ noise and signal samples, respectively. This means that 1 in 4 candidates with this ranking statistic value would be astrophysical in origin. Following the same procedure, our second-most significant candidate would have a 1 in 25 chance of being astrophysical in origin.

An examination of the \textit{Fermi}-GBM counts in the twelve individual NaI detectors does not reveal a single detector primarily responsible for the observed excess. A time-frequency representation of the LIGO data is shown in Fig~\ref{fig:data}. While no strong transient noise similar to that which occurred during GW170817 is evident, we cannot rule out the possibility of some non-stationary noise affecting the data during this time.

We estimate posterior probabilities of the event's GW parameters by performing a Bayesian analysis on the LIGO data with PyCBC Inference \citep{Biwer:2018osg, pycbc-github} and using the TaylorF2 post-Newtonian waveform \citep{Sathyaprakash:1991mt,Droz:1999qx,Blanchet:2002av,Faye:2012we} to model the gravitational wave. For this analysis we use a uniform prior in component masses $m_{1,2} \in [1,3)\,\msun$, which spans the entire range of known neutron-star masses \citep{Ozel:2016oaf}. The dimensionless spins of each component $\chi_{1,2}$ are constrained to be aligned with the orbital-angular momentum with a prior that is uniform in $[-0.4, 0.4)$. This choice is consistent with the fastest-known spinning neutron star \citep{Hessels:2006ze}. The tidal deformability parameters $\Lambda_{1,2}$ of the components are varied independently of each other, with a prior uniform in $[0, 5000)$. We use a prior isotropic in binary orientation and uniform in volume between $5$ and $500\,$Mpc. A uniform prior spanning $t_c \pm 0.1\,$s is used for the coalescence time, where $t_c$ is the coalescence time estimated by the search. To measure likelihood, $128\,$s of data spanning $t_c-110\,s$ to $t_c+18\,$s are filtered between $20\,$ and $2048\,$Hz. The power spectral density of each detector is estimated using Welch's method with $512\,$s of data centered on the event. No marginalization over calibration uncertainty is performed.

\begin{figure}
    \centering
    \includegraphics[width=\columnwidth]{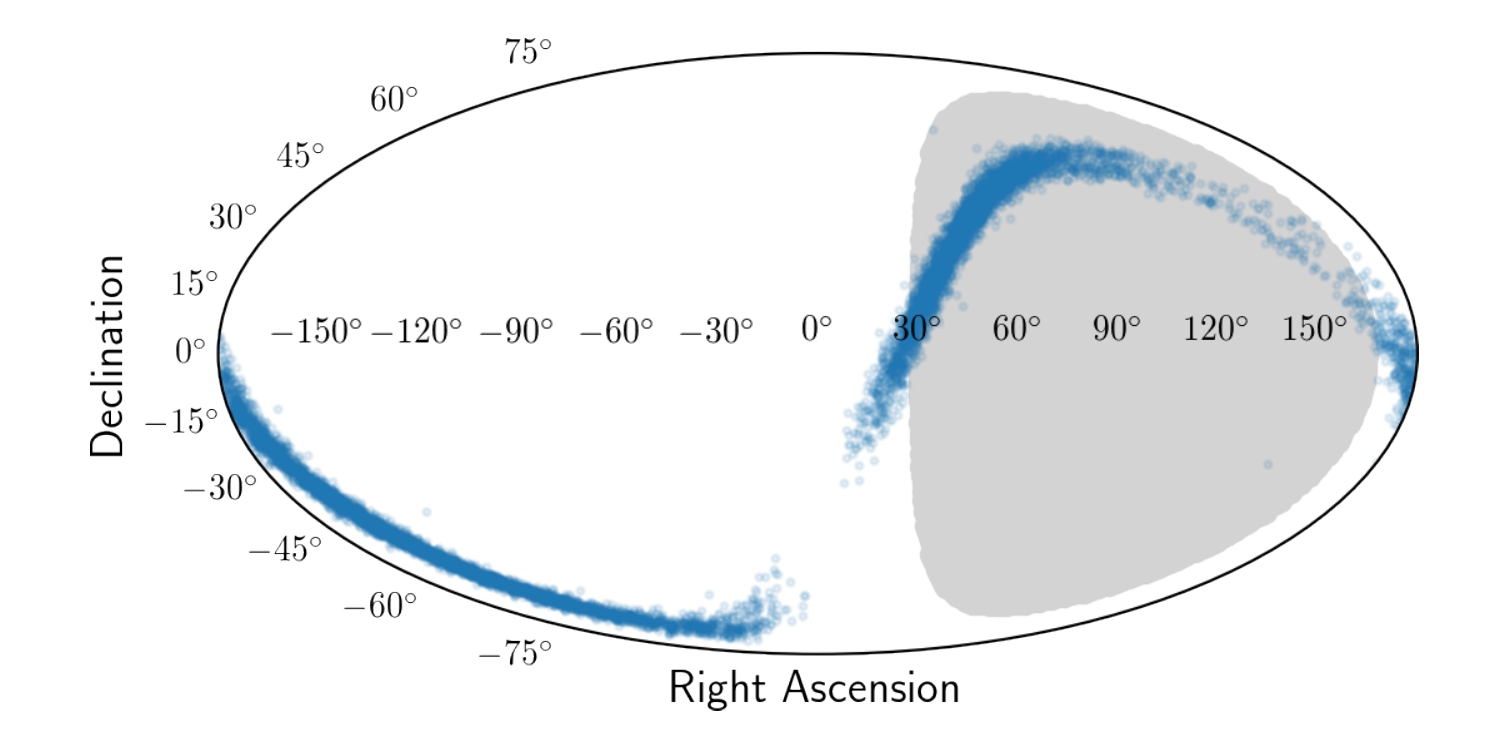}
    \caption{Sky localization for 1-OGC 151030. Samples from the posterior probability are shown in blue. The gray
    region shows the portion of the sky occulted by the Earth from the perspective of the \textit{Fermi} satellite at the time of the event. If 1-OGC 151030 is a real GW-GRB observation then it cannot have come from the $40\%$ of the GW sky localization that is blocked by the Earth. This corresponds to $B_{loc} \sim 0.8$}.
    \label{fig:skymap}
\end{figure}

The results of the parameter estimation are shown in Figs.~\ref{fig:posteriors} and \ref{fig:skymap}. Figure \ref{fig:posteriors} shows the marginal posterior distributions of the source-frame chirp mass $\mathcal{M}^{\mathrm{src}}$, mass ratio $q$, effective spin $\chieff$, inclination $\iota$ and luminosity distance $d_L$. The median and 90\% credible intervals for each parameter are quoted above the 1D marginal plots, with the exception of the inclination angle (due to the bimodal posterior) and the mass ratio (since it is largely unconstrained). To estimate $\mathcal{M}^{\mathrm{src}}$ from the observed, detector-frame chirp mass, we assume a standard $\Lambda$CDM cosmology \citep{Ade:2015xua}. Figure~\ref{fig:skymap} shows a 2D marginal distribution of the sky location. The tidal parameters $\Lambda_{1,2}$ (not shown) are not constrained, which is expected for low SNR sources.

 We also perform parameter estimation separately on data from each GW detector. The single-detector SNRs recovered by this analysis are consistent with the coherent SNR of the signal. However, the detector-frame chirp mass posterior shows multiple peaks in the Livingston detector. This may indicate the presence of non-stationary or non-Gaussian noise in the Livingston detector. To determine if such peaks can be observed with a known signal in similar data, we repeated the analysis on a simulated signal with similar parameters added to the data, offset by $t_c+0.157\,$s. Since we find similar peaks in the recovered chirp mass of the simulated signal, the 1-OGC 151030 Livingston results do not necessarily rule out an astrophysical source.

The degeneracy between inclination angle and luminosity distance is evident in the right plot of Fig.~\ref{fig:posteriors}. The bimodal posterior of the inclination angle is typical when the data is uninformative about the viewing angle. This is expected for BNSs in which no independent redshift information is available~\citep{Chen:2018omi}. The degeneracy leads to a larger uncertainty in the luminosity distance. However, if we assume that the viewing angle is $< 30^\circ$ in order to be detected by \textit{Fermi}-GBM, then we obtain a luminosity distance of $\measconedist\,$Mpc.

By their very nature, sub-threshold detections will dominant the population of the most distant signals. As such, these sub-threshold events will always provide the strongest constraints on the speed of gravity. In our search we have explicitly assumed that a GRB trigger should lie within 3.4 seconds of a gravitational-wave trigger in order to be coincident. Assuming, however, that 1-OGC 151030 is a real astrophysical event seen in both gravitational waves and gamma rays, we can use the arrival times to constrain the speed of gravity relative to the speed of light, $\delta v = v_{GW} - v_{EM} $. Following the method of \cite{Monitor:2017mdv} and using a 90\% lower bound on the distance of 116 Mpc, we obtain
\begin{equation}
-6 \times 10^{-16}  \leq \frac{\delta v}{v_{EM}} \leq 3 \times 10^{-16} .
\end{equation}
The upper (positive) bound is a factor 2 stronger than \cite{Monitor:2017mdv} and the lower (negative) bound is a factor 5 stronger. This improvement is mainly due to the larger luminosity distance of the signal compared to GW170817.

\section{Discussion}
We search for BNS mergers that are observable both by their GW and associated GRB emission. We have combined the likelihood that these multi-messenger candidates are associated with a true GW signal and the likelihood of sky localization agreement with a GRB source. With future observations, this methodology could be further improved by including the likelihood that a GRB candidate is astrophysical for sub-threshold GRB events along with a model that relates the GW and GRB observables, such as GW observed masses and GRB energy spectrum.

A major uncertainty in the sensitivity of this analysis is the unknown
selection bias of the GRB candidates. If this population of GRB candidates is just as likely to contain the counterpart to a sub-threshold GW event as a clearly detected GW, then the overall sensitive distance is 20\% greater, corresponding to 70\% in volume, than a gravitational-wave search alone. This indicates that approximately 40\% of all GW-GRB events that could be observed will only be found by this kind of search. However, if our population of GRBs does not contain the possible counterparts to quiet GW events, then the practical sensitivity of this search will be lower than expected. Given the rate of short GRBs, it is beneficial to examine lower threshold candidates which cannot necessarily be discerned as GRBs on their own. We have shown that setting a threshold such that we recover candidates at a false alarm rate of 1 per 3 hours still allows the search to reach our target sensitivity. Methods discussed in~\cite{Blackburn:2014rqa,Goldstein:2016zfh} demonstrate that coherently combining the Fermi-GBM data can increase sensitivity to weak sources. Future work may significantly improve the purity of our GRB sample and include more detailed information which can be correlated with gravitational-wave observations.

The main advantage of a joint analysis is the reduction in non-astrophysical background compared to a GW or GRB search alone. This background reduction primarily arises from the requirement that a GW and GRB candidate occur in close temporal proximity. Our analysis also takes into account the agreement between GW and GRB localizations when available. However, this is not a strict constraint due to the large uncertainty in source location for most of our sample. Improved localization for the majority of our candidates would further reduce the background. In the case of 1-OGC 151030, the original false alarm rate of this candidate from the 1-OGC analysis was 1 per 2 hours~\citep{Nitz:2018imz}. If we restrict our analysis to just BNS-like sources, the false alarm rate would have been 1 per 2 days. By combining GW-only analysis with information from \textit{Fermi}-GBM, this candidate was promoted to 1 per 13 years. Improved localization of the GRB component of this candidate may be able to further strengthen or weaken the association between the GW and GRB observations.

Even if the event 1-OGC 151030 is not a true GW-GRB observation, the prospects for detecting such signals in the near future in data from the second or third LIGO observing run seem compatible with more optimistic scenarios~\citep{Howell:2018nhu}. In the years to come, the combination of information from different astronomical observations will be of increasing importance and will likely include not just gravitational-wave and gamma-ray observations, but also other parts of the electromagnetic spectrum, neutrinos and cosmic rays~\citep{Branchesi:2016vef}. In particular, surveys of kilonova candidates with the Large Synoptic Survey Telescope may provide another population which could be further correlated with GW candidates and increase the reach of multi-messenger searches~\citep{Andreoni:2018fcm,Setzer:2018ppg}.

To aid follow-up, we make available supplementary materials which include sky localizations for our GW candidates and the posterior samples for 1-OGC 151030~\citep{o1-gwgrb}.

\acknowledgments
We thank Sylvia Zhu, Eric Burns, Sebastian Khan,  Duncan Brown, Tito Dal Canton, and Thomas Dent for their feedback.
We acknowledge the Max Planck Gesellschaft and the Atlas cluster computing team at AEI Hanover for support. This research has made use of data, software and/or web tools obtained from the Gravitational Wave Open Science Center (https://www.gw-openscience.org), a service of LIGO Laboratory, the LIGO Scientific Collaboration and the Virgo Collaboration. LIGO is funded by the U.S. National Science Foundation. Virgo is funded by the French Centre National de Recherche Scientifique (CNRS), the Italian Istituto Nazionale della Fisica Nucleare (INFN) and the Dutch Nikhef, with contributions by Polish and Hungarian institutes.
\bibliography{references}

\end{document}